\documentclass[fleqn,10pt,preprint,1p]{elsarticle}
\usepackage[intlimits,reqno]{amsmath}
\usepackage{cite}
\biboptions{sort&compress}
\allowdisplaybreaks
\def\oo{\infty}
\def\Z#1{\zeta(#1)}
\def\A#1{a_{#1}}
\def\B#1{b_{#1}}
\def\D#1{d_{#1}}
\def\Re{\mathrm{Re}}
\def\Im{\mathrm{Im}}

\def\appzero{{\left(\frac{\alpha_0}{\pi}\right)}}
\def\appzeromu{{\left(\frac{\alpha_0(\mu)}{\pi}\right)}}
\def\appstrong{{\left(\frac{\alpha_s(\mu)}{\pi}\right)}}
\def\mumq{{\left(\frac{\mu^2}{m^2}\right)}}

\def\nlep{n_h}
\def\Li{{\mathrm{Li}}}
\def\cl#1#2{{\mathrm{Cl}_{#1}}\left({#2}\right)}
\def\Cl#1{{\mathrm{Cl}_{#1}\left(\frac{\pi}{3}\right)}}
\def\Cldd{{\mathrm{Cl^2_2}\left(\frac{\pi}{3}\right)}}
\def\cat#1{{\mathrm{Cl}_{#1}\left(\frac{\pi}{2}\right)}}
\def\catdd{{\mathrm{Cl}^2_{2}\left(\frac{\pi}{2}\right)}}
\def\ZTD{\zeta^2(3)}
\def\rha#1{\Re H_{#1}\left(e^{i\frac{\pi}{3}}\right)}
\def\rhb#1{\Re H_{#1}\left(e^{i\frac{2\pi}{3}}\right)}
\def\iha#1{\Im H_{#1}\left(e^{i\frac{\pi}{3}}\right)}
\def\ihb#1{\Im H_{#1}\left(e^{i\frac{2\pi}{3}}\right)}
\def\rhe#1{\Re H_{#1}\left(e^{i\frac{\pi}{2}}\right)}
\def\ihe#1{\Im H_{#1}\left(e^{i\frac{\pi}{2}}\right)}
\def\rt{\sqrt{3}}
\def\rs{\sqrt{s}}
\def\btell{B_3}
\def\ctell{C_3}
\def\WF{{\tilde{F}}}
\def\intaxoneones#1{f_1(#1)}
\def\intaxonetwos#1{f_2(#1)}

\def\e{\epsilon}

\def\TITLE{High-precision four-loop mass and wave function renormalization in QED}

\def\CAPTIONFIGA{A selection of some of 4-loop self-mass diagrams,
with the indication of the part of \eqref{zsplit} to which the diagram contributes.
The numbering of the diagrams follows Ref.\citep{Laporta:2017okg,Laporta:2019fmy}.}

\def\eqref#1{Eq.(\ref{#1})}
\def\eqrefb#1#2{Eqs.(\ref{#1})-(\ref{#2})}

\def\comba#1{{t_#1}}
\def\combp#1{{v_#1}}
\def\combel#1{{e_#1}}
\def\TableZd 
{
\begin{table}\centering
\scriptsize
\begin{center}
\begin{tabular}{cccccc}
\hline
\\[-6pt]
 & $\e^{-4}$
 & $\e^{-3}$
 & $\e^{-2}$
 & $\e^{-1}$
 & $\e^{0}$ \\
\hline
$\delta Z_2^{(4)}$
&    0.205023871527777 
&    0.597746672453703 
&  - 0.893282495748014 
&  - 6.188211339005751
&  - 17.26913874640770 \\
$\delta Z_2^{(4)}$\citep{Marquard:2018rwx} 
& $  0.20500(37)$
& $  0.5980(27)$
& $ -0.895(21)$
& $ -6.18(17)$
& $ -17.4(16)$ \\
\hline
$\delta Z_2^{(4,0)}$
&   0.01318359375
&   0.08349609375
& - 0.082767885375100
& - 1.965268322191886 
& - 4.036104196851619 \\
$\delta Z_2^{FFFF}$\citep{Marquard:2018rwx} 
&   $0.01317(25)$
&   $0.0836(19)$
&  $-0.084(71)$
&  $-1.96(16)$
&  $-4.1(15)$ \\
\hline
$\delta Z_2^{(4,l-l)}$
& 0
& 0
& 0
& - 0.125
&   0.105307603143802 \\
$\delta Z_2^{dFFH}$\citep{Marquard:2018rwx} 
& $-0.00001(23)$
& $ 0.0001(15)$
& $-0.001(11)$
& $-0.120(76)$
& $ 0.10(50)$\\
\hline
$\delta Z_2^{(4,1)}$
&   0.0703125
&   0.255859375
& - 0.655004826193796
& - 3.800454407485363
& - 5.953609261982407 \\
$\delta Z_2^{FFFH}$\citep{Marquard:2018rwx}
& $ 0.070313(23)$
& $ 0.255860(93)$
& $-0.65497(55)$
& $-3.8002(36)$
& $-5.953(19)$ \\
\hline
$\delta Z_2^{(4,2)}$
&    0.09375
&    0.2109375 
&  - 0.329091743327423 
&  - 0.574390474672734
&  - 7.996856441249995 \\
$\delta Z_2^{FFHH}$\citep{Marquard:2018rwx}
& $ 0.0937498(14)$
& $ 0.2109378(59)$
& $-0.329095(35)$
& $-0.57438(13)$
& $-7.99681(79)$ \\
\hline
$\delta Z_2^{(4,3)}$
& 0.027777777777777 
& 0.047453703703703 
& 0.173581959148306 
& 0.276901865344232 
& 0.612123550532518 \\
$\delta Z_2^{FHHH}$\citep{Marquard:2018rwx}
&  0.0277778
&  0.047454
&  0.173582
&  0.276902
&  0.61212 \\
\hline
\end{tabular}
\end{center}
\caption{Comparison between our values of $\delta Z_2$ and the
QED and QCD results of Ref.\citep{Marquard:2018rwx}.} 
\label{TableZdlab}
\end{table}
}
\def\Figuraboh 
{ 
\begin{figure} 
\begin{center}
\includegraphics[scale=0.875]{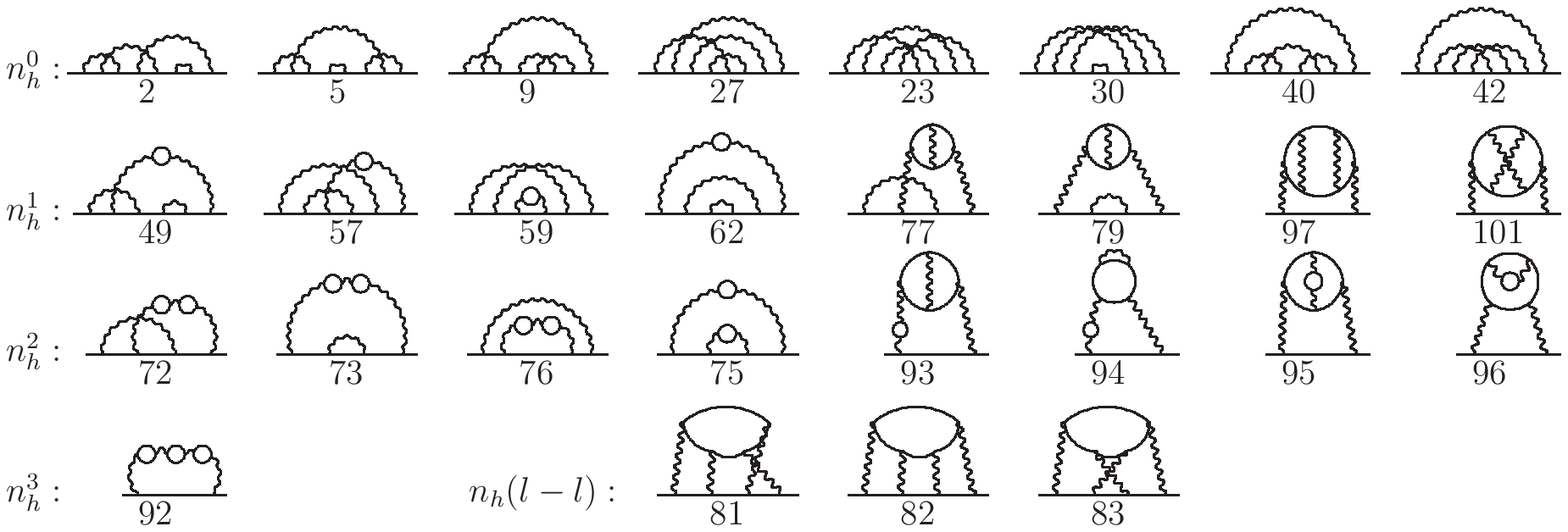}
\caption{\CAPTIONFIGA}
\label{self104}
\end{center}
 \end{figure} 
 }            
\begin{document}
\begin{frontmatter}
%%%%%%%%%%%%%%%%%%%%%%%%%%%%%%%%%%%%%%%%%%
%%%%%%%%%%%%%%%%%%%%%%%%%%%%%%%%%%%%%%%%%%
% for article'
\title{\TITLE }
\author{Stefano Laporta}
\ead{stefano.laporta@pd.infn.it}
\address{Dipartimento di Fisica e Astronomia, Universit\`a di Padova,
Istituto Nazionale Fisica Nucleare, Sezione di Padova,
Via Marzolo 8, I-35131 Padova, Italy}
%%%%%%%%%%%%%%%%%%%%%%%%%%%%%%%%%%%%%%%%%%
%%%%%%%%%%%%%%%%%%%%%%%%%%%%%%%%%%%%%%%%%%
\begin{abstract}
The 4-loop QED mass and wave function renormalization constants
$Z_2$ and $Z_m$ have been evaluated in the on-shell subtraction scheme 
with 1100 digits of precision. 
We also worked out the coefficients of the five color structures of 
the QCD renormalization constants $Z_2^{\rm OS}$ and $Z_m^{\rm OS}$ which can
obtained from QED-like diagrams.
The results agree with lower precision results available in the literature.
Analytical fits were also obtained for all these quantities.
\end{abstract}
\end{frontmatter}
%%%%%%%%%%%%%%%%%%%%%%%%%%%%%%%%%%%%%%%%%%
%%%%%%%%%%%%%%%%%%%%%%%%%%%%%%%%%%%%%%%%%%
%%%%%%%%%%%%%%%%%%%%%%%%%%%%%%%%%%%%%%%%%%%
\section{Introduction}
The 4-loop QED contribution to the electron $g$-$2$  was
calculated with 1100 digits of precision in Ref.\citep{Laporta:2017okg}.
In that paper high-precision numerical and analytical fits were also obtained 
for all the master integrals of the 4-loop self-mass QED diagrams.
Therefore, other 4-loop quantities that can be expressed in terms of the same
master integrals can be known with such precision. 
In Ref.\citep{Laporta:2019fmy}
we used these results 
to obtain
a high-precision value and 
an analytical fit of   
the 4-loop first derivative of the Dirac form factor $F_1'(0)$.

In this third paper, following the same approach as in
Ref.\citep{Laporta:2017okg,Laporta:2019fmy}, we calculate
high-precision numerical values and analytical expression 
of the 4-loop QED wave function and mass renormalization
constants $Z_2$ and $Z_m$,
in the on-shell subtraction scheme.
We infer also the values of the coefficients of five color structures
of the QCD renormalization constants $Z_2^{\rm OS}$ and $Z_m^{\rm OS}$ which
can be obtained from the QED result.

\section{Definitions}
The wave function renormalization constant $Z_2$ is defined as
as $\psi_0=\sqrt{Z_2} \psi$,  where $\psi_0$ and $\psi$ are the
bare and the renormalized electron field, respectively.
The mass renormalization constant $Z_m$ is defined as $m_0=Z_m m $, where $m_0$
and $m$ are the bare and the physical electron mass, respectively.
%QED renormalization constants can be extract from the QCD result.
Two-loop QED and QCD analytical results for on-shell $Z_2$ and $Z_m$ 
were obtained in Ref.\citep{Broadhurst:1991fy}; three-loop 
analytical results were obtained in Refs.\citep{Melnikov:2000zc,Marquard:2007uj}.

In QCD $m_0=Z_m^{\rm OS} m^{\rm OS}=Z_m^{\overline{\rm MS}} m^{\overline{\rm
MS}}$. Being $Z_m^{\overline{\rm MS}}$ known to sufficiently
high degree of perturbative expansion\citep{Chetyrkin:1999pq,Luthe:2016xec,Baikov:2017ujl}, 
the ratio $m^{\rm OS}/m^{\overline{\rm MS}}$ can be used to determine
the on-shell $Z_m^{\rm OS}$; 
the ratio $m^{\rm OS}/m^{\overline{\rm MS}}$
%it 
is known at two loops
analytically\citep{Gray:1990yh}, and at three loops it was obtained  
in numerical form in Ref.\citep{Chetyrkin:1999ys,Chetyrkin:1999qi}
and in analytical form in Ref.\citep{Melnikov:2000qh,Marquard:2007uj}.
%The electromagnetic field renormalization constant is defined as 
%$A_{\mu}^0 = \sqrt{Z_3} A_{mu}$, 
%where $A_{\mu}^0$ and $A_{\mu}$ 
%are the bare and renormalized electromagnetic field.
%The charge renormalization constant $Z_1$ is 
%defined as 
%$e_0 = \frac{Z_1}{Z_2\sqrt{Z_3}} $ where $e_0$ and $e$ are 
%the bare and renormalized coupling constant;
%is not necessary to calculated it because $Z_1=Z_2$ as consequence
%of the Ward identity; anyway we will use this as a check. 

Four-loop QED and QCD numerical results of  $Z_2$ were obtained 
in Ref.\citep{Lee:2013sx,Marquard:2018rwx,Marquard:2018qqx};
%, with worst error up to $36$\%.
the four-loop numerical result of $Z_m$ can be worked out from the QCD
results for $z_m$ and $Z_m^{\rm OS}$ of 
Ref.\citep{Lee:2013sx,Marquard:2015qpa,Marquard:2016dcn};
see also Ref.\citep{Kataev:2015gvt,Kataev:2018aaa,Kataev:2018aab}.
Both results have a precision
of a few digits; see section~\ref{Comparisons} for more details.
%%, with worst error up to $10$\%.
%Both papers calculated some master integrals with MonteCarlo methods.
%The four-loop analytical expression of $Z_3$ was obtained in Ref.\citep{Lee:2013sx}.

%for both ultraviolet and infrared divergences;
$Z_2$ and $Z_m$ are gauge parameter independent 
(for $Z_2$ see Ref.\citep{Landau:1955zz,Johnson:1959zz,Fukuda:1978jy});
in order to simplify the calculations, we choose
the Feynman gauge. 

%\begin{equation}
%(-(n-1)psl p_{\mu}-(n-1)*I*m*p_{\mu})/(4n)/m^2*(-i)^N
%\end{equation}

%%%%%%%%%%%%%%%%%%%%%%%%%%%%%%%%%%%%%%%%%%
%%%%%%%%%%%%%%%%%%%%%%%%%%%%%%%%%%%%%%%%%%%%%%%%%%%%%%%%%%%%%%%%%%%%%%%%%%%%%%
%We follow the normalization used in Ref.\citep{Melnikov:2000zc}
%for the 3-loop results of $Z_2$ and $Z_m$.
We expand $Z_2$ and $Z_m$ in power series of the bare coupling constant $\alpha_0$:
\begin{equation}\label{z2exp1}
Z_2=1+\sum_{i} \appzero^i
\left(\Gamma(1+\e)(4\pi)^{\e} m^{-2\e}\right)^i
Z_2^{(i)} \ , 
%\end{equation}
%\begin{equation}\label{zmexp1}
\qquad
Z_m=1+\sum_{i} \appzero^i
\left(\Gamma(1+\e)(4\pi)^{\e} m^{-2\e}\right)^i
Z_m^{(i)} \ .
\end{equation}

We decompose the 4-loop coefficients $Z_2^{(4)}$ and $Z_m^{(4)}$ 
by considering QED with $\nlep$ different leptons of mass $m$;
in this way we separate the contributions coming from diagrams with different
number of insertions of vacuum polarizations and light-light
scattering diagrams. 
We use the notation $\nlep$ in order to adapt to the usual QCD convention
of calling the number of massive fermions $n_h$
and the number of massless fermions $n_l$. 
In this paper we do not consider diagrams with massless leptons.
Therefore
\begin{align}\label{zsplit}
Z_2^{(4)}(\nlep)&=
Z_2^{(4,0)}
+\nlep   Z_2^{(4,1)}
+\nlep^2 Z_2^{(4,2)}
+\nlep^3 Z_2^{(4,3)}
+\nlep   Z_2^{(4,l-l)}
\ ,
\cr
Z_m^{(4)}(\nlep)&=
Z_m^{(4,0)}
+\nlep   Z_m^{(4,1)}
+\nlep^2 Z_m^{(4,2)}
+\nlep^3 Z_m^{(4,3)}
+\nlep   Z_m^{(4,l-l)}
\ .
\end{align}
Clearly 
\begin{align}\label{zsplit2}
Z_2^{(4)}(\nlep=1)=& Z_2^{(4)}
\ ,
%\cr
\qquad
Z_m^{(4)}(\nlep=1)= Z_m^{(4)}
\  .
\end{align}
In Fig.\ref{self104}
we show a selection of some of the
4-loop self-mass diagrams, with the indication of the part of
\eqref{zsplit} to which
the diagram contributes. The complete set of 104 diagrams
is shown in Ref.\citep{Laporta:2017okg,Laporta:2019fmy}.
\Figuraboh

\subsection{Method}
We briefly describe here the method used to obtain our results.
At 4-loop level there are 104 QED self-mass diagrams. 
The contribution to $Z_2^{(4)}$ and $Z_m^{(4)}$ from each self-mass diagram is
extracted by using projectors and taking traces 
with a {\tt FORM}\citep{FORM,Kuipers:2012rf}
program; it is a linear combination of Feynman integrals.
Then, for each self-mass diagram a system of integration-by-parts (I.B.P.)
identities\citep{Chetyrkin:1981qh,Tkachov:1981wb}
is build and solved\citep{Laporta:2001dd}
in order to reduce the Feynman integrals to linear combinations 
of master integrals. 
We used the systems of I.B.P. identities generated in
Ref.\citep{Laporta:2017okg} for the 4-loop $g$-$2$.
Counterterms were added where needed, 
excluding vacuum polarizations, since we expand in the bare coupling
constant $\alpha_0$.
The renormalization constants are reduced 
to linear combinations of the 4-loop $g$-$2$
master integrals:  
\begin{equation}\label{relaz}
Z_2^{(4)}=\sum_{i=1}^{N} C_{2,i}(\e) M_i(\e)\ , 
\qquad \qquad
Z_m^{(4)}=\sum_{i=1}^{N} C_{m,i}(\e) M_i(\e)\ ,
\end{equation}
where $C_{2,i}$ and $C_{m,i}$ are rational functions in $\e$, $N=334$.
We use the numerical values and analytical fits of the master integrals $M_i$
worked out in Ref.\citep{Laporta:2017okg}.
As a simple consistency check, we extracted $Z_1^{(4)}$ from the 
891 4-loop vertex diagrams, and we checked numerically and analytically 
the Ward identity  $Z_1^{(4)}=Z_2^{(4)}$.

\section{Results}
\subsection{Numerical results}
By substituting the numerical values of $M_i(\e)$ in \eqref{relaz},
we have obtained 1100-digits numerical values for 
$Z_2^{(4,x)}$
and
$Z_m^{(4,x)}$. 
We show here results truncated to 40 digits for the sake of space.
Full-precision results are available from the author.
The numerical value of the coefficients of the powers of $\nlep$ 
in \eqref{zsplit} are:

\begin{align}\label{z2piece}
   Z_2^{(4,0)} =&\;
            0.01318359375 \e^{-4}
          + 0.08349609375 \e^{-3}
          - 0.1261401703408252765467615072316270440815 \e^{-2}
	  \cr&\;
          - 2.218829553807290472156162364826322487455  \e^{-1}
	  \cr&\;
          - 3.572910387812835300654933535440420039948   +O(\e)\ ,
\\
   Z_2^{(4,1)}=&\;
            0.0703125 \e^{-4} 
          + 0.255859375 \e^{-3}
          - 0.8863236793443281079737642171893946198198 \e^{-2}
	  \cr&\;
          - 4.529505177334142374400047352337054775442  \e^{-1}
	  \cr&\;
          - 3.084249643446330546559819112760212296306   +O(\e)\ ,
\\
   Z_2^{(4,2)}=&\;
            0.09375 \e^{-4}
          + 0.2109375 \e^{-3}
          - 0.6375168808614659083936734252065216879251 \e^{-2}
	  \cr&\;
          - 1.118089921229380504506879422676331183925  \e^{-1}
	  \cr&\;
          - 6.170238285159180066980183430726643994320   +O(\e)\ ,
\\
   Z_2^{(4,3)}=&\;
            0.02777777777777777777777777777777777777777 \e^{-4}
	  \cr&\;
          + 0.04745370370370370370370370370370370370370 \e^{-3}
	  \cr&\;
          + 0.08219673321229348679616766668042765163181 \e^{-2}
	  \cr&\;
          + 0.1653060637464923649096813755300430397436  \e^{-1}
	  \cr&\;
          + 0.2373760011149585346199605381719332090858    +O(\e)\ ,
\\
   Z_2^{(4,l-l)}=&\; - 0.125 \e^{-1}
                   + 0.1053076031438024383339691487092823499443  +O(\e)\ ,
\end{align}
\begin{align}\label{zmpiece}
   Z_m^{(4,0)}=&\;
            0.01318359375 \e^{-4}
          + 0.05712890625 \e^{-3}
          + 0.2149248550926856846841693020449834317223 \e^{-2}
	  \cr&\;
          - 0.6311216133387257157783705667342740710866 \e^{-1} 
	  \cr&\;
          - 6.640775996670789293945443649244052753483  +O(\e)\ ,
\\
   Z_m^{(4,1)}=&\;
            0.03515625 \e^{-4}
          + 0.0171875 \e^{-3}
          + 0.2255475542195506536638492553197210259614 \e^{-2}
	  \cr&\;
          - 3.655124674567472080082766471095472176020  \e^{-1}
	  \cr&\;
          - 1.052445900250388864170227694635348345572  +O(\e)\ ,
\\
   Z_m^{(4,2)}=&\;
            0.02864583333333333333333333333333333333333 \e^{-4}
          + 0.171875 \e^{-3}
	  \cr&\;
          + 0.02125506230176173571324393225701890093679 \e^{-2}
	  \cr&\;
          - 1.803410367796313396577956494832032506831 \e^{-1}
	  \cr&\;
          + 1.933766152908452159688148893799688168422  +O(\e)\ ,
\\
   Z_m^{(4,3)}=&\;
            0.006944444444444444444444444444444444444444 \e^{-4}
	  \cr&\;
          + 0.05439814814814814814814814814814814814814 \e^{-3}
	  \cr&\;
          + 0.002035866606146743398083833340213825815905 \e^{-2}
	  \cr&\;
          + 0.4351198769462674456781824350266903384281 \e^{-1}
	  \cr&\;
          - 0.8481242413000520746932873847887178671417  +O(\e)\ ,
\\
    Z_m^{(4,l-l)}=&\;
            0.5009641733560598212811272632084921831710 \e^{-1}
	  \cr&\;
          - 3.947552272425901748117750000830748333155  +O(\e)\ .
\end{align}	  
%are available from Ref.\citep{Laporta:2019okg}. 
From \eqref{zsplit2} we obtain the values of the renormalization
constants:

\begin{align}
Z_2^{(4)} =   &\; 0.2050238715277777777777777777777777777777 \e^{-4}\cr
       &\;+ 0.5977466724537037037037037037037037037037 \e^{-3}\cr
       &\;- 1.567783997334325806118031482947115700194 \e^{-2}\cr
       &\;- 7.826118588624320986153407764309665407079 \e^{-1}\cr
       &\;- 12.48471471215958494124100639204606077154 +O(\e)
\ ,
\label{z2num}
\end{align}
\begin{align}\label{zmnum}
Z_m^{(4)}=    &\; 0.08393012152777777777777777777777777777777\e^{-4}\cr
       &\;+ 0.4552770543981481481481481481481481481481\e^{-3}\cr
       &\;+ 0.4637633382201448174593463229619371844364\e^{-2}\cr
       &\;- 5.153572605400183925479783834426596232339\e^{-1}\cr
       &\;- 10.55513225773867982123855983569917913093 +O(\e)
\ .
\end{align}

%%%%%%%%%%%%%%%%%%%%%%%%%%%%%%%%%%%%%%%%%%%%%%%%%%%%%%%%%%%%%%%%%
%%%%%%%%%%%%%%%%%%%%%%%%%%%%%%%%%%%%%%%%%%%%%%%%%%%%%%%%%%%%%%%%%
%%%%%%%%%%%%%%%%%%%%%%%%%%%%%%%%%%%%%%%%%%%%%%%%%%%%%%%%%%%%%%%%%
\subsection{Comparisons}
\label{Comparisons}
\subsubsection{QED wave function renormalization constant}
Now we compare our results 
\eqrefb{z2piece}{zmpiece}
and
\eqrefb{z2num}{zmnum}
with the numerical results of
Ref.\citep{Marquard:2018rwx};
In Ref.\citep{Marquard:2018rwx} $Z_2$ is written as:
%, taking into account the different
%normalization.
%Therefore we re-write \eqrefb{z2exp1}{zmexp1} :
\begin{equation}
Z_2=
1+\sum_{i} \appzeromu^i
\left( {(4\pi)^{\e}}{e^{-\gamma \e}\mumq^{\e}}\right)^i
\delta Z_2^{(i)} \ .
\end{equation}
The relation between $\delta Z_2^{(4)}$ and our $Z_2^{(4)}$ for $\mu=1$ is
\begin{equation}
\delta Z_2^{(4)}= \left(\frac{e^{-\gamma\e}}{\Gamma(1+\e)}\right)^4
Z_2^{(4)}
\ .
\end{equation}
%%%%%%%%%%%%%%%%%%%%%%%%%%%%%%%%%%%%%%%%%%%%%%%%%%%%%%%%%%%%%%%%%%%%%%%%
In the first row of Table \ref{TableZdlab} we show our 
high-precision numerical values of the coefficients of the powers of
$\e$ of  $\delta Z_2^{(4)}$, truncated to 15 digits for reason of space,
and the corresponding values from Ref.\citep{Marquard:2018rwx}.
They are in good agreement, the worst error being $0.1\sigma$.
In the next subsection we will need to decompose $\delta Z_2$ 
in terms with different $\nlep$,
\begin{equation}\label{z24split}
\delta Z_2^{(4)}=
\delta Z_2^{(4,0)}
+\nlep   \delta Z_2^{(4,1)}
+\nlep^2 \delta Z_2^{(4,2)}
+\nlep^3 \delta Z_2^{(4,3)}
+\nlep   \delta Z_2^{(4,l-l)}\ ;
\end{equation}
our numerical values of $\delta Z_2^{(4)}$ are listed in Table
\ref{TableZdlab};
preliminar values were presented in Ref.\citep{Laporta:2018eos}.
%%%%%%%%%%%%%%%%%%%%%%%%%%%%%%%%%%%%%%%%%%%%%%%%%%%%%%%%%%%%
\subsubsection{QCD wave function renormalization constant}
We consider now the QCD
renormalization constant $Z_2^{QCD}$ in the OS renormalization scheme.
In the notation of Ref.\citep{Marquard:2018rwx}
$Z_2^{QCD}$ is decomposed in 23 color structures:
%we explicitly write here only the structures whose coefficients 
%can obtained from QED-like diagrams.
%containing only $C_F$ and massive fermions, 
%[meglio dire fundamental repre]
\begin{equation}
\delta Z_2^{QCD} = 
C_F^4 \delta  Z_2^{FFFF}+ 
C_F^3 T n_h  \delta Z_2^{FFFH}+ 
C_F^2 T^2 n_h^2 \delta Z_2^{FFHH}+ 
C_F T^3 n_h^3 \delta Z_2^{FHHH}+ 
n_h \frac{d_F^{abcd} d_F^{abcd}}{N_c} \delta Z_2^{d_{FF}H} +{\text{\ldots} } \ .
\end{equation}
The coefficients $Z_2^{(4,x)}$ of \eqref{z24split}
must coincide with the corresponding coefficients of the color structures
which can be obtained from  QED-like diagrams:
%%calculated in the Feynman gauge:
\begin{align}
& \delta Z_2^{4,0}=\delta Z_2^{FFFF} \ ,  
\quad
  \delta Z_2^{4,1}=\delta Z_2^{FFFH} \ ,  
\quad
  \delta Z_2^{4,2}=\delta Z_2^{FFHH} \ ,  
& \delta Z_2^{4,3}=\delta Z_2^{FHHH} \ ,  
\quad
  \delta Z_2^{4,l-l}=\delta Z_2^{d_{FF}H} \ .  
\end{align}
\TableZd
In Table \ref{TableZdlab} we compare our high-precision results
and the corresponding ones from Ref.\citep{Marquard:2018rwx}.
They are in good agreement, the worst error being $0.08\sigma$.

%%%%%%%%%%%%%%%%%%%%%%%%%%%%%%%%%%%%%%%%%%%%%%%%%%%%%%%%%%%%%%
\subsubsection{QCD mass renormalization constant}
%%\citep{Marquard:2015qpa,Marquard:2016dcn}

Now we consider the ratio between the QCD mass renormalization constants
in the ${\rm OS}$ and $\overline{\rm MS}$ scheme:
\begin{equation}
z_m(\mu) = \frac{Z_m^{\rm OS}}{Z_m^{\overline{\rm MS}}}=1+\sum_{n\ge
1}\appstrong^n z_m^{(n)}(\mu) 
\ ;
\end{equation}
%[dove e' $Z_m^{MS}$?]
Using the notation of Ref.\citep{Marquard:2016dcn}, 
$z_m$ is decomposed in 23 color structures, and as above we can infer the
coefficients of the five structures which involve only QED-like diagrams. 
\begin{equation}
z_m^{(4)}= 
C_F^4 z_m^{FFFF}+ 
C_F^3 T n_h  z_m^{FFFH}+ 
C_F^2 T^2 n_h^2 z_m^{FFHH}+ 
C_F T^3 n_h^3 z_m^{FHHH}+ 
n_h \frac{d_F^{abcd} d_F^{abcd}}{N_c} z_m^{d_{FF}H} +{\ldots} 
\end{equation}
Our high-precision values are 
\begin{align}
z_m^{FFFF}&=    - 6.943004942063674366729783085730127607323  \ ,\cr
z_m^{FFFH}&=    - 1.364670155556599206268818327382565081078  \ ,\cr
z_m^{FFHH}&= \phantom{+}  1.657513434712808758859954030175433075697  \ ,\\
z_m^{FHHH}&=    - 0.1490239616711777449135125845999523299403 \ ,\cr
z_m^{d_{FF}H}&= - 3.947552272425901748117750000830748333155  \ .
\nonumber
\end{align}
The results of Ref.\citep{Marquard:2016dcn} 
\begin{align}
z_m^{FFFF}&=   -6.983(805)  \ ,\cr
z_m^{FFFH}&=   -1.3625(132) \ , \cr
z_m^{FFHH}&= \phantom{+}  1.65752(31)  \ ,\\
z_m^{FHHH}&=   -0.14902     \ ,\cr
z_m^{d_{FF}H}&= -3.924(642) 
\ ,
\nonumber
\end{align}
are in agreement with ours at the level of $0.16 \sigma$ at worst.
%%%%%%%%%%%%%%%%%%%%%%%%%%%%%%%%%%%%%%%%%%%%%%%%%%%%%%%%%%%
%%%%%%%%%%%%%%%%%%%%%%%%%%%%%%%%%%%%%%%%%%%%%%%%%%%%%%%%%%%%%%%%%%%%
%%%%%%%%%%%%%%%%%%%%%%%%%%%%%%%%%%%%%%%%%%%%%%%%%%%%%%%%%%%%%%%%%%%%
%%%%%%%%%%%%%%%%%%%%%%%%%%%%%%%%%%%%%%%%%%%%%%%%%%%%%%%%%%%
\subsection{Analytical fits}
By substituting the analytical fits of $M_i(\e)$ 
in \eqref{relaz},
we have obtained
the following analytical expressions:
\begin{align}\label{z1anat}
Z_2^{(4)}&=
           \frac{3779}{18432 \e^4}
          +\frac{33053}{55296 \e^3}
          +\frac{1}{\e^2}\biggl(
	   \frac{515315}{73728}
          -\frac{7205}{768}\Z2 
          -\frac{131}{64}\Z3
          +\frac{131}{16}\Z2 \ln 2
          \biggr)
          +\frac{1}{\e} \biggl(
           \frac{19571293}{663552}
\cr&
          +\frac{154747}{2160}\Z2
          -\frac{11521}{2304}\Z3
          -\frac{29539}{192}\Z2 \ln 2
          +\frac{3115}{64}\Z4
          -\frac{215}{8}\Z2 \ln^2 2
          -\frac{19}{4} \comba{4}
          -\frac{5}{64}\Z5
          -\frac{21}{16}\Z3\Z2
          \biggr)
\cr&
          +\frac{9565004502941}{87787929600}
          +\frac{1535743349}{691200}\Z2
          +\frac{12128503957}{25401600}\Z3
          +\frac{10500647}{17280}\Z2 \ln 2
\cr&
          -\frac{535034261}{1451520}\Z4
          +\frac{5631023}{9072}\Z2 \ln^2 2
          +\frac{7932313}{9072} \comba{4}
          -\frac{2050259}{17280}\Z5
\cr&
          +\frac{9296423}{8640}\Z3\Z2
          -\frac{609737}{480}\Z4 \ln 2
          +\frac{1739}{24}\Z2 \ln^3 2
          -\frac{2671}{30} \comba{5}
          -\frac{715229459}{62208}\Z6
\cr&
          -\frac{4006421}{11520}\ZTD
          +\frac{1780957}{240}\Z3\Z2 \ln 2
          -\frac{4689809}{720}\Z4 \ln^2 2
          -\frac{9674}{45}   \comba{{61}}
          +\frac{10276}{45}  \comba{{62}}
          +\frac{642767}{90} \comba{{63}}
\cr&
          +\frac{1495323863}{580608}\Z7
          -\frac{5775661427}{161280}\Z5\Z2
          -\frac{7455877}{10752}\Z4\Z3
          +\frac{2153119}{128}\Z6 \ln 2
\cr&
          -\frac{561331}{126}\Z3\Z2 \ln^2 2
          -\frac{19454}{63}  \comba{{71}}
          +\frac{9144}{7}    \comba{{72}}
          +\frac{4306}{7}    \comba{{4}}\Z3
          +\frac{553037}{15} \comba{{5}}\Z2
          +\frac{135039}{5}  \comba{{73}}
\cr&
+\rt\biggl(
           \frac{15489}{320}  \Cl4
          -\frac{1311089}{960} \Z2 \Cl2
          +\frac{3071}{60}       \combp{{61}}
          +\frac{2109}{50}       \combp{{62}}
          +\frac{7472227}{34320} \combp{{63}}
          -\frac{8978057}{6480}  \combp{{64}}
	  \biggr)
\cr&
          +\frac{5797}{16} \combp{{65}}
          +\frac{115735}{96}\Z2 \Cldd
	  +18 \combp{{71}}
          +6 \combp{{72}}
          -36\Z2 \cat2
          -\frac{12606}{5}\Z2 \cat2^2
\cr&
      +\rt\pi\biggl(
          -\frac{10163659}{230400}  \btell
          +\frac{224075873}{6220800} \ctell
          -\frac{11863}{7776} \intaxonetwos{0,0,1}
          +\frac{56207}{23328} \combel{{51}}
          -\frac{14615}{1728}  \combel{{52}}
          -\frac{45499}{20736} \combel{{61}}
\cr&
          +\frac{30961}{41472} \combel{{62}}
          \biggr)
   + \Z2 \biggl(
          -\frac{2354}{243}    \intaxoneones{0,0,1}
          -\frac{30961}{3456}  \combel{{53}}
          -\frac{673}{162}     \combel{{54}}
	  \biggr)
\cr&
          - \frac{507}{80} C_{81a}
          -26 C_{81b}
          +\frac{11}{2} C_{81c}
          -\frac{1057}{320} C_{83a}
          +\frac{91}{24} C_{83b}
          +\frac{7}{2} C_{83c} 
	   +O(\e)
	  \ ,
\end{align}
%%%%%%%%%%%%%%%%%%%%%%%%%%%%%%%%%%%%%%%%%%%%%%%%%%%%%%%%%%%%%%%%%%%%
%%%%%%%%%%%%%%%%%%%%%%%%%%%%%%%%%%%%%%%%%%%%%%%%%%%%%%%%%%%%%%%%%%%%
\begin{align}\label{zmanat}
Z_m^{(4)}&=
           \frac{1547}{18432 \e^4}
          +\frac{25175}{55296 \e^3}
          +\frac{1}{\e^2}
	  \biggl(
           \frac{202951}{73728}
          -\frac{2737}{768}\Z2
          -\frac{71}{128}\Z3
          +\frac{119}{32}\Z2 \ln 2
         \biggr)
	  +\frac{1}{\e}
	  \biggl(
           \frac{27750271}{1990656}
\cr&
          +\frac{132763}{5760}\Z2
          -\frac{1625}{768}\Z3
          -\frac{10807}{192}\Z2 \ln 2
          +\frac{6173}{384}\Z4
          -\frac{553}{48}\Z2 \ln^2 2
          +\frac{35}{24} \comba{4}
          -\frac{5}{8}\Z5
\cr&
          +\frac{21}{32}\Z3\Z2
          \biggr)
          +\frac{1885204711}{23887872}
          +\frac{5804091169}{6220800}\Z2
          +\frac{169571}{768}\Z3
          +\frac{87397}{5760}\Z2 \ln 2
          -\frac{1866527}{6912}\Z4
\cr&
          +\frac{74477}{288}\Z2 \ln^2 2
          +\frac{62645}{144} \comba{4}
          -\frac{15459}{256}\Z5
          +\frac{20351}{48}\Z3\Z2
          -\frac{33367}{64}\Z4 \ln 2
          +\frac{357}{16}\Z2 \ln^3 2
\cr&
          +\frac{35}{4} \comba{5}
          -\frac{8229601}{1728}\Z6
          -\frac{5605}{32}\ZTD
          +\frac{12297}{4}\Z3\Z2 \ln 2
          -\frac{10975}{4}\Z4 \ln^2 2
          +3040 \comba{{63}}
          -\frac{161}{128}\Z7
\cr&
          -\frac{27786101}{1920}\Z5\Z2
          -\frac{272627}{384}\Z4\Z3
          +\frac{252105}{32}\Z6 \ln 2
          -\frac{3675}{2}\Z3\Z2 \ln^2 2
          +\frac{74368}{5} \comba{{5}}\Z2
\cr&
          +\frac{53368}{5} \comba{{73}}
+\rt\biggl(
           45 \Cl4
          -\frac{12757}{24} \Z2 \Cl2
          +\frac{98}{5}   \combp{{62}}
          -\frac{4949}{9} \combp{{64}}
	  \biggr)
          +180\combp{{65}}
\cr&
          +\frac{993}{2}\Z2 \Cldd
          - 1008\Z2 \cat2^2
	 +\rt\pi\biggl(	  
          -\frac{754571}{46080} \btell
          +\frac{17787301}{1244160} \ctell
          -\frac{503}{1728} \intaxonetwos{0,0,1}
          +\frac{895}{324}  \combel{{51}}
\cr&
          -\frac{1295}{216} \combel{{52}}
          -\frac{1493}{1728}\combel{{61}}
          +\frac{671}{3456} \combel{{62}}
	  \biggr)
   + \Z2 \biggl(
          -\frac{100}{27}  \intaxoneones{0,0,1}
          -\frac{671}{288} \combel{{53}}
          -\frac{346}{27}  \combel{{54}}
	  \biggr)
          - \frac{21}{8} C_{81a}
          -10 C_{81b}
\cr&
          -\frac{21}{16} C_{83a}
          +\frac{5}{3} C_{83b} 
	   +O(\e)
	  \ .
\end{align}
The analytical expressions of the separate contributions of 
\eqref{zsplit} are:
\begin{align}
   Z_2^{(4,0)} &=
           \frac{27}{2048 \e^4}
          +\frac{171}{2048 \e^3}
          +\frac{1}{\e^2}\biggl(
           \frac{5835}{8192}
          -\frac{351}{256}\Z2
          -\frac{27}{64}\Z3
          +\frac{27}{16}\Z2\ln 2
          \biggr)
       +\frac{1}{\e}\biggl(
           \frac{23865}{8192}
          +\frac{4171}{128}\Z2
\cr&	  
          +\frac{2527}{256}\Z3
          -\frac{4713}{64}\Z2\ln 2
          +\frac{909}{64}\Z4
          -\frac{81}{8}\Z2\ln^2 2
          +\frac{57}{4}\comba{{4}}
          -\frac{25}{64}\Z5
          -\frac{9}{16}\Z3\Z2
          \biggr)
          +\frac{2033213}{98304}
\cr&	  
          +\frac{6966313}{15360}\Z2
          +\frac{352489}{11520}\Z3
          +\frac{642189}{640}\Z2\ln 2
          +\frac{9612919}{23040}\Z4
          +\frac{574357}{1440}\Z2\ln^2 2
          +\frac{2341}{720}\comba{{4}}
\cr&	  
          -\frac{1149787}{5760}\Z5
          +\frac{108635}{192}\Z3\Z2
          -\frac{83371}{160}\Z4\ln 2
          +\frac{161}{8}\Z2\ln^3 2
          +\frac{1727}{10}\comba{{5}}
          -\frac{38991047}{5184}\Z6
\cr&	  
          -\frac{756779}{11520}\ZTD
          +\frac{414137}{120}\Z3\Z2\ln 2
          -\frac{1227247}{360}\Z4\ln^2 2
          -\frac{19994}{45}\comba{{61}}
          +\frac{25156}{45}\comba{{62}}
          +\frac{24104}{9}\comba{{63}}
\cr&	  
          -\frac{9389399}{82944}\Z7
          -\frac{8309201}{512}\Z5\Z2
          +\frac{4441247}{4608}\Z4\Z3
          +\frac{1096627}{128}\Z6\ln 2
          -\frac{72457}{36}\Z3\Z2\ln^2 2
\cr&	  
          -\frac{136}{9}\comba{{71}}
          -\frac{32}{3}\comba{{72}}
          -\frac{1}{3}\comba{{4}}\Z3
          +\frac{50705}{3}\comba{{5}}\Z2
          +{12331}\comba{{73}}
       +\rt\biggl(
          -\frac{6949}{192}\Cl4
          -\frac{563899}{576}\Z2\Cl2
\cr&	  
          +\frac{3071}{60}\combp{{61}}
          +\frac{2109}{50}\combp{{62}}
          +\frac{7472227}{34320}\combp{{63}}
          -\frac{8978057}{6480}\combp{{64}}
          \biggr)
          +\frac{26125}{32}\Z2\Cldd
          -\frac{1995}{16}\combp{{65}}
          -{18}\combp{{71}}
          -{6}\combp{{72}}
\cr&	  
       +\rt\pi\biggl(
          -\frac{32021}{25600} \btell
          +\frac{3656149}{230400}\ctell
          -\frac{4775}{1296}\intaxonetwos{0,0,1}
          -\frac{965143}{23328}\combel{{51}}
          +\frac{32885}{576}\combel{{52}}
          -\frac{10705}{6912}\combel{{61}}
          +\frac{585}{512}\combel{{62}}
          \biggr)
\cr&	  
          +\Z2\biggl(
          -\frac{2414}{243}\intaxoneones{0,0,1}
          -\frac{1755}{128}\combel{{53}}
          +\frac{3565}{54}\combel{{54}}
          \biggr)
	   +O(\e)
	  \ ,
\end{align}
%===========================================================================
\begin{align}
 Z_2^{(4,l)} &=
          -\frac{1}{8 \e}
          -\frac{205}{128}
          +\frac{2563}{2}\Z2
          +\frac{35933}{192}\Z3
          +\frac{1419}{2}\Z2\ln 2
          -\frac{15155}{192}\Z4
          -\frac{679}{3}\Z2\ln^2 2
          +\frac{628}{3}\comba{{4}}
          -\frac{51259}{192}\Z5
\cr&	  
          +\frac{95239}{120}\Z3\Z2
          -\frac{20975}{16}\Z4\ln 2
          -{16}\comba{{5}}
          -\frac{56584517}{10368}\Z6
          -\frac{541547}{1920}\ZTD
          +\frac{46507}{10}\Z3\Z2\ln 2
\cr&	  
          -\frac{20947}{6}\Z4\ln^2 2
          +\frac{688}{3}\comba{{61}}
          -\frac{992}{3}\comba{{62}}
          +\frac{81632}{15}\comba{{63}}
          +\frac{195131207}{72576}\Z7
          -\frac{394782889}{20160}\Z5\Z2
\cr&	  
          -\frac{6682045}{4032}\Z4\Z3
          +\frac{264123}{32}\Z6\ln 2
          -\frac{615463}{252}\Z3\Z2\ln^2 2
          -\frac{18502}{63}\comba{{71}}
          +\frac{27656}{21}\comba{{72}}
          +\frac{12925}{21}\comba{{4}}\Z3
\cr&	  
          +\frac{299512}{15}\comba{{5}}\Z2
          +\frac{73384}{5}\comba{{73}}
          +\rt\biggl(
           \frac{693}{8}\Cl4
          -\frac{2523}{8}\Z2\Cl2
          \biggr)
          +\frac{4969}{16}\Z2\Cldd
          +\frac{3873}{8}\combp{{65}}
\cr&	  
          +{36}\combp{{71}}
          +{12}\combp{{72}}
          -{36}\Z2\cat2
          -\frac{12606}{5}\Z2\catdd
        +\rt\pi\biggl(
          -\frac{15399}{320}\btell
          +\frac{1477}{192} \ctell
          +\frac{857}{288}\intaxonetwos{0,0,1}
\cr&	  
          +\frac{39785}{972}\combel{{51}}
          -\frac{13235}{216}\combel{{52}}
          -\frac{1673}{2592}\combel{{61}}
          -\frac{2053}{5184}\combel{{62}}
          \biggr)
       +\Z2\biggl(
           \frac{20}{81}\intaxoneones{0,0,1}
          +\frac{2053}{432}\combel{{53}}
          -\frac{5684}{81} \combel{{54}}
	  \biggr) 
\cr&	  
          -\frac{507}{80}C_{81a}
          -{26}C_{81b}
          +\frac{11}{2}C_{81c}
          -\frac{1057}{320}C_{83a}
          +\frac{91}{24}C_{83b}
          +\frac{7}{2}C_{83c}
	   +O(\e)
	  \ ,
\end{align}
%===========================================================================
\begin{align}
Z_2^{(4,1)}&=
           \frac{9}{128 \e^4}
          +\frac{131}{512 \e^3}
       +\frac{1}{\e^2 }\biggl(
           \frac{1747}{768}
          -\frac{135}{32}\Z2
          -\frac{9}{8}\Z3
          +\frac{9}{2}\Z2\ln 2
          \biggr)
       +\frac{1}{\e}\biggl(
           \frac{142385}{18432}
          +\frac{50989}{1152}\Z2
\cr&	  
          +\frac{2777}{768}\Z3
          -\frac{747}{8}\Z2\ln 2
          +\frac{891}{32}\Z4
          -\frac{63}{4}\Z2\ln^2 2
          +\comba{{4}}
          +\frac{5}{16}\Z5
          -\frac{3}{4}\Z3\Z2
          \biggr)
          +\frac{934395461}{37324800}
\cr&	  
          +\frac{8610967}{19440}\Z2
          +\frac{809922361}{3110400}\Z3
          -\frac{394219}{432}\Z2\ln 2
          -\frac{24640541}{34560}\Z4
          +\frac{57655}{144}\Z2\ln^2 2
          +\frac{26603}{36}\comba{{4}}
\cr&	  
          +\frac{450661}{1440}\Z5
          -\frac{8861}{27}\Z3\Z2
          +\frac{136613}{240}\Z4\ln 2
          +\frac{121}{3}\Z2\ln^3 2
          -\frac{1886}{15}\comba{{5}}
          +\frac{92170207}{62208}\Z6
\cr&	  
          -\frac{1}{32}\ZTD
          -\frac{10899}{16}\Z3\Z2\ln 2
          +\frac{6185}{16}\Z4\ln^2 2
          -\frac{1957}{2}\comba{{63}}
       +\rt\biggl(
          -\frac{487}{240}\Cl4
\cr&	  
          -\frac{51373}{720}\Z2\Cl2
	  \biggr)
          +\frac{3773}{48}\Z2\Cldd
          +\frac{23}{8}\combp{{65}}
       +\rt\pi\biggl(
           \frac{237613}{34560}   \btell
          +\frac{6777463}{933120} \ctell
\cr&	  
          -\frac{1733}{1944}\intaxonetwos{0,0,1}
          +\frac{3695}{1296}\combel{{51}}
          -\frac{3695}{864}\combel{{52}}
          \biggr)
	   +O(\e)
	  \ ,
\end{align}
%===========================================================================
\begin{align}
Z_2^{(4,2)}&=
           \frac{3}{32 \e^4}
          +\frac{27}{128 \e^3}
        +\frac{1}{\e^2} \biggl(
           \frac{4337}{1536}
          -\frac{25}{8}\Z2
          -\frac{1}{2}\Z3
          +{2}\Z2\ln 2
          \biggr)
       +\frac{1}{\e} \biggl(
           \frac{167545}{13824}
          -\frac{3827}{2160}\Z2
\cr&	  
          -\frac{31843}{2304}\Z3
          +\frac{55}{6}\Z2\ln 2
          +\frac{53}{8}\Z4
          -\Z2\ln^2 2
          -{20}\comba{{4}}
          \biggr)
          +\frac{40705891003}{1828915200}
          +\frac{3264023}{48600}\Z2
\cr&	  
          +\frac{3697035983}{152409600}\Z3
          -\frac{2133}{10}\Z2\ln 2
          -\frac{5625569}{362880}\Z4
          +\frac{4638449}{90720}\Z2\ln^2 2
          -\frac{964289}{22680}\comba{{4}}
          +\frac{1889}{54}\Z5
\cr&	  
          +\frac{1609}{36}\Z3\Z2
          -\frac{15}{2}\Z4\ln 2
          +{12}\Z2\ln^3 2
          -{120}\comba{{5}}
       +\rt\pi\biggl(
          -\frac{111683}{69120}\btell
          +\frac{9696589}{1866240}\ctell
\cr&	  
          +\frac{145}{1944}\intaxonetwos{0,0,1}
          \biggr)
	   +O(\e)
	  \ ,
\end{align}
%===========================================================================
\begin{align}
Z_2^{(4,3)}&=
           \frac{1}{36 \e^4}
          +\frac{41}{864 \e^3}
	  +\frac{1}{\e^2}\biggl(
           \frac{679}{576}
          -\frac{2}{3}\Z2
	  \biggr)
       +\frac{1}{\e}\biggl(
           \frac{71143}{10368}
          -\frac{103}{30}\Z2
          -\frac{14}{3}\Z3
          +{4}\Z2\ln 2
          \biggr)
\cr&	  
          +\frac{166888903}{3919104}
          -\frac{31451}{1350}\Z2
          -\frac{37691}{1512}\Z3
          +\frac{103}{5}\Z2\ln 2
          +\frac{259}{12}\Z4
          -\frac{10}{3}\Z2\ln^2 2
          -\frac{104}{3}\comba{{4}}
	   +O(\e)
	  \ ,
\end{align}
%===========================================================================
%===========================================================================
\begin{align}
   Z_m^{(4,0)} &=
           \frac{27}{2048 \e^4}
          +\frac{117}{2048 \e^3}
          +\frac{1}{\e^2}\biggl(
           \frac{3063}{8192}
          -\frac{135}{256}\Z2
          -\frac{27}{128}\Z3
          +\frac{27}{32}\Z2\ln 2
          \biggr)
       +\frac{1}{\e} \biggl(
           \frac{28653}{8192}
          +\frac{1479}{128}\Z2
\cr&	  
          +\frac{471}{256}\Z3
          -\frac{1773}{64}\Z2\ln 2
          +\frac{747}{128}\Z4
          -\frac{63}{16}\Z2\ln^2 2
          +\frac{45}{8}\comba{{4}}
          -\frac{15}{32}\Z5
          +\frac{9}{32}\Z3\Z2
          \biggr)
          +\frac{2907301}{98304}
\cr&	  
          +\frac{166719}{1024}\Z2
          +\frac{1109}{128}\Z3
          +\frac{56537}{128}\Z2\ln 2
          +\frac{7613}{128}\Z4
          +\frac{4149}{32}\Z2\ln^2 2
          +\frac{765}{16}\comba{{4}}
          -\frac{11123}{256}\Z5
\cr&	  
          +\frac{3429}{16}\Z3\Z2
          -\frac{13867}{64}\Z4\ln 2
          +\frac{81}{16}\Z2\ln^3 2
          +\frac{135}{4}\comba{{5}}
          -\frac{22615}{8}\Z6
          +\frac{219}{32}\ZTD
          +{1235}\Z3\Z2\ln 2
\cr&	  
          -{1390}\Z4\ln^2 2
          +{1018}\comba{{63}}
          -\frac{189}{64}\Z7
          +\frac{110845}{256}\Z4\Z3
          -\frac{1658509}{256}\Z5\Z2
          +\frac{108045}{32}\Z6\ln 2
\cr&	  
          -\frac{1575}{2}\Z3\Z2\ln^2 2
          +{6698}\comba{{5}}\Z2
          +{4898}\comba{{73}}
          +\rt\biggl(
          -\frac{2983}{8}\Z2\Cl2
          +\frac{98}{5}  \combp{{62}}
          -\frac{4949}{9}\combp{{64}}
	  \biggr)
\cr&	  
          +\frac{675}{2}\Z2\Cldd
          +\rt\pi\biggl(
          -\frac{3123}{5120} \btell
          +\frac{33383}{5120}\ctell
          -\frac{781}{576}\intaxonetwos{0,0,1}
          -\frac{5245}{324}\combel{{51}}
          +\frac{1615}{72} \combel{{52}}
\cr&	  
          -\frac{721}{1152} \combel{{61}}
          +\frac{1075}{2304}\combel{{62}}
          \biggr)
	  +\Z2\biggl(
          -\frac{100}{27}\intaxoneones{0,0,1}
          -\frac{1075}{192}\combel{{53}}
          +\frac{245}{9}\combel{{54}}
          \biggr)
	   +O(\e)
	  \ ,
\end{align}
%===========================================================================
\begin{align}
   Z_m^{(4,l)} &=
        \frac{1}{\e}\biggl(
          -\frac{1}{16}
          +\frac{15}{32}\Z3
          \biggr)
          -\frac{135}{128}
          +{555}\Z2
          +\frac{18625}{192}\Z3
          +{48}\Z2\ln 2
          -\frac{803}{64}\Z4
          -{20}\Z2\ln^2 2
          +{80}\comba{{4}}
\cr&	  
          -\frac{7205}{96}\Z5
          +\frac{1273}{4}\Z3\Z2
          -{603}\Z4\ln 2
          -\frac{2134769}{864}\Z6
          -\frac{1453}{8}\ZTD
          +{2086}\Z3\Z2\ln 2
          +{2384}\comba{{63}}
\cr&	  
          -{1490}\Z4\ln^2 2
          +\frac{217}{128}\Z7
          -\frac{877789}{768}\Z4\Z3
          -\frac{30694567}{3840}\Z5\Z2
          +\frac{36015}{8}\Z6\ln 2
          -{1050}\Z3\Z2\ln^2 2
\cr&	  
          +\frac{40878}{5}\comba{{5}}\Z2
          +\frac{28878}{5}\comba{{73}}
       +\rt\biggl(
           {45}\Cl4
          -{129}\Z2\Cl2
          \biggr)
          +{126}\Z2\Cldd
          +{180}\combp{{65}}
\cr&	  
          -{1008}\Z2\catdd
       +\rt\pi\biggl(
          -\frac{2859}{160}\btell
          +\frac{1253}{480}\ctell
          +\frac{299}{216}\intaxonetwos{0,0,1}
          +\frac{1450}{81}\combel{{51}}
          -\frac{725}{27} \combel{{52}}
          -\frac{823}{3456} \combel{{61}}
\cr&	  
          -\frac{1883}{6912}\combel{{62}}
          \biggr)
       +\Z2\biggl(
           \frac{1883}{576}\combel{{53}}
          -\frac{1081}{27} \combel{{54}}
          \biggr)
          -\frac{21}{8}C_{81a}
          -{10} C_{81b}
          -\frac{21}{16}C_{83a}
          +\frac{5}{3}C_{83b}
	   +O(\e)
	  \ ,
\end{align}
%===========================================================================
\begin{align}
   Z_m^{(4,1)} &=
           \frac{9}{256 \e^4}
          +\frac{11}{64 \e^3}
          +\frac{1}{\e^2}\biggl(
           \frac{2507}{3072}
          -\frac{93}{64}\Z2
          -\frac{9}{32}\Z3
          +\frac{15}{8}\Z2\ln 2
          \biggr)
          +\frac{1}{\e}\biggl(
           \frac{4363}{1024}
          +\frac{91}{6}\Z2
\cr&	  
          +\frac{51}{128}\Z3
          -{35}\Z2\ln 2
          +\frac{71}{8}\Z4
          -\frac{25}{4}\Z2\ln^2 2
          +\frac{5}{2}\comba{{4}}
          -\frac{5}{32}\Z5
          +\frac{3}{8}\Z3\Z2
	  \biggr)
          +\frac{1273135}{36864}
\cr&	  
          +\frac{1312775}{6912}\Z2
          +\frac{87181}{768}\Z3
          -\frac{3187}{8}\Z2\ln 2
          -\frac{2063917}{6912}\Z4
          +{133}\Z2\ln^2 2
          +{320}\comba{{4}}
\cr&	  
          +\frac{1567}{32}\Z5
          -\frac{257}{2}\Z3\Z2
          +\frac{4653}{16}\Z4\ln 2
          +\frac{45}{4}\Z2\ln^3 2
          +{15}\comba{{5}}
          +\frac{34251}{64}\Z6
\cr&	  
          -\frac{3}{8}\ZTD
          -\frac{987}{4}\Z3\Z2\ln 2
          +\frac{545}{4}\Z4\ln^2 2
          -{362}\comba{{63}}
          -\frac{89}{3}\rt\Z2\Cl2
          +{33}\Z2\Cldd
\cr&	  
          +\rt\pi\biggl(
           \frac{3599}{1280}\btell
          +\frac{97783}{34560}\ctell
          -\frac{227}{648}\intaxonetwos{0,0,1}
          +\frac{85}{81}\combel{{51}}
          -\frac{85}{54}\combel{{52}}
          \biggr)
	   +O(\e)
	  \ ,
\end{align}
%===========================================================================
\begin{align}
   Z_m^{(4,2)} &=
           \frac{11}{384 \e^4}
          +\frac{11}{64 \e^3}
       +\frac{1}{\e^2}\biggl(
           \frac{1555}{1536}
          -\frac{5}{4}\Z2
          -\frac{1}{16}\Z3
          + \Z2\ln 2
	  \biggr)
        +\frac{1}{\e}\biggl(
           \frac{3343}{1152}
          -\frac{329}{180}\Z2
          -\frac{97}{32}\Z3
\cr&	  
          +\frac{53}{12}\Z2\ln 2
          +\frac{131}{96}\Z4
          -\frac{4}{3}\Z2\ln^2 2
          -\frac{20}{3}\comba{{4}}
          \biggr)
          -\frac{183577}{55296}
          +\frac{1777157}{48600}\Z2
          +\frac{7043}{576}\Z3
\cr&	  
          -\frac{31397}{360}\Z2\ln 2
          -\frac{44239}{1728}\Z4
          +\frac{335}{18}\Z2\ln^2 2
          +\frac{5}{9}\comba{{4}}
          +\frac{439}{48}\Z5
          +\frac{239}{12}\Z3\Z2
\cr&	  
          +\frac{15}{2}\Z4\ln 2
          +{6}\Z2\ln^3 2
          -{40}\comba{{5}}
          +\rt\pi\biggl(
          -\frac{8159}{11520}\btell
          +\frac{726817}{311040}\ctell
          +\frac{5}{162}\intaxonetwos{0,0,1}
          \biggr)
	   +O(\e)
	  \ ,
\end{align}
%===========================================================================
\begin{align}
   Z_m^{(4,3)} &=
           \frac{1}{144 \e^4}
          +\frac{47}{864 \e^3}
       +\frac{1}{\e^2}\biggl(
           \frac{317}{576}
          -\frac{1}{3}\Z2
          \biggr)
       +\frac{1}{\e}\biggl(
           \frac{103963}{31104}
          -\frac{83}{45}\Z2
          -\frac{43}{24}\Z3
          +{2}\Z2\ln 2
          \biggr)
\cr&	  
          +\frac{3579989}{186624}
          -\frac{7622}{675}\Z2
          -\frac{1529}{144}\Z3
          +\frac{166}{15}\Z2\ln 2
          +\frac{347}{48}\Z4
          -\frac{8}{3}\Z2\ln^2 2
          -\frac{40}{3}\comba{{4}}
	   +O(\e)
	  \ .
\end{align}

%%%%%%%%%%%%%%%%%%%%%%%%%%%%%%%%%%%%%%%%%%%%%%%%%%%%%%%%%%%%%%%%%%%%
%%%%%%%%%%%%%%%%%%%%%%%%%%%%%%%%%%%%%%%%%%%%%%%%%%%%%%%%%%%%%%%%%%%%
%%%%%%%%%%%%%%%%%%%%%%%%%%%%%%%%%%%%%%%%%%%%%%%%%%%%%%%%%%%%%%%%%%%%
%%%%%%%%%%%%%%%%%%%%%%%%%%%%%%%%%%%%%%%%%%%%%%%%%%%%%%%%%%%%%%%%%%%%
%%%%%%%%%%%%%%%%%%%%%%%%%%%%%%%%%%%%%%%%%%%%%%%%%%%%%%%%%%%%%%%%%%%%%%%%%%%%%%
In the above expressions we use these combinations of constants:
\begin{align}\label{taa}
\comba{4}=&\A4+\frac{1}{24}\ln^4 2 \ ,\qquad
\comba{5}=\A5  + \frac{1}{12}\Z2\ln^3 2 -\frac{1}{120}\ln^5 2 \ ,
  \\
\comba{{61}}=&\B6- \A5 \ln 2 + \Z5 \ln 2 + \frac{1}{6}\Z3 \ln^3 2 -\frac{1}{12}\Z2\ln^4 2 + \frac{1}{144}\ln^6 2
\ ,
\\
\comba{{62}}=&\A6 -\frac{1}{48}\Z2 \ln^4 2 +\frac{1}{720}\ln^6 2
\ ,
\qquad \comba{{63}}=\comba{4} \Z2 \ ,
\end{align}
\begin{align}
\comba{{71}}=& \D7 - 2\B6 \ln 2 + 4\A6\ln 2 + 2 \A5 \ln^2 2 -
\frac{49}{32}\ZTD \ln 2 -\frac{95}{32} \Z5 \ln^2 2
       +\frac{1}{8} \Z4  \ln^3 2
       \cr&
       -\frac{1}{3} \Z3  \ln^4 2
       +\frac{1}{12} \Z2  \ln^5 2 -\frac{1}{120} \ln^7 2 
	  \ ,
\end{align}
\begin{align}
\comba{{72}}=&\B7
 - 3 \A7 - \A6 \ln 2 -\frac{1}{2} \Z5  \ln^2 2 +\frac{1}{48} \Z4  \ln^3 2 -
 \frac{1}{24} \Z3 \ln^4 2 
%\cr&
 +\frac{ 1}{120} \Z2 \ln^5 2
 -\frac{1}{1680} \ln^7 2 \ ,
%%% -\frac{ \ln^7 2}{1680} \ ,
\end{align}
\begin{align}
\comba{{73}}=&\left(
\A4-\frac{1}{4}\Z2 \ln^2 2  +\frac{ 7}{16} \Z3 \ln 2 +\frac{ 1}{24}
\ln^4 2
\right) \Z2 \ln 2 
\ ,
\end{align}
\begin{align}
\combp{{61}}=& \iha{0,0,0,1,-1,-1} 
       + \ihb{0,0,0,1,-1,1}
       + \ihb{0,0,0,1,1,-1}
       + \frac{27}{26}\ihb{0,0,1,0,1,1}
\cr&
       + \frac{207}{104}\ihb{0,0,0,1,1,1}
       + \frac{10}{3}\A4 \Cl2
       + \frac{7}{4} \Z3 \iha{0,1,-1}
       + \frac{21}{8} \Z3 \ihb{0,1,1}
         \cr&
       - \frac{5}{72} \Z3 \Z2 \pi
       - \frac{5}{6} \Cl2 \Z2\ln^2 2
       + \frac{5}{36} \Cl2 \ln^4 2
       - \frac{27413}{67392} \Z5 \pi
       + \frac{4975}{11583} \Z4 \Cl2
	  \ ,
\end{align}
\begin{align}
\combp{{62}}& = \Z2 \biggl(
         \iha{0,1,1,-1} 
       + \frac{3}{2} \ihb{0,1,1,-1}
       - \frac{1}{6} \Z3 \pi
       + \frac{1}{108} \Z2 \pi \ln 2
       - \frac{5}{2} \iha{0,1,-1} \ln 2
       \cr&
       - \frac{15}{4} \ihb{0,1,1} \ln 2
       + \frac{25}{12} \Cl2 \ln^2 2
%       \cr&
       - \frac{661}{1188} \Cl2 \Z2
       \biggr)
	  \ ,
\end{align}
\begin{align}
\combp{{63}}=&\Cl6 - \frac{3}{4} \Z4 \Cl2 \ , \qquad      
\combp{{64}} = \Cl4 \Z2 - \frac{91}{66} \Z4 \Cl2  \ ,
\\
\combp{{65}}=&\rha{0,0,0,1,0,1}+\Cl2 \Cl4\ ,
\end{align}
\begin{align}
\combp{{71}}=&\rha{0,0,0,1,0,1,-1} 
       + 4 \rha{0,0,0,0,1,1,-1}
       - \frac{27}{8} \rhb{0,0,1,0,0,1,1}
\cr&
       - \frac{135}{16} \rhb{0,0,0,1,0,1,1}
       - \frac{27}{2} \rhb{0,0,0,0,1,1,1}
\cr&
       + \iha{0,1,-1} \Cl4
       + \frac{3}{2} \ihb{0,1,1} \Cl4
       + \frac{145}{132} \Cl6 \pi 
       \ ,
\end{align}
\begin{align}
\combp{{72}}=& \Z2 \biggl(
         \rha{0,1,0,1,-1}
       + 2 \rha{0,0,1,1,-1}
       + \frac{9}{4} \rhb{0,1,0,1,1}
       + \frac{9}{2} \rhb{0,0,1,1,1}
\cr&
       + \iha{0,1,-1} \Cl2
       + \frac{3}{2} \ihb{0,1,1} \Cl2
       \biggr)
       \ ,
\end{align}
\begin{align}
\combp{{73}}=&\Z2 \biggl(
      \rhe{0,1,0,1,1}
       + \cat2 \ihe{0,1,1}
       - \frac{1}{2} \cat4 \pi
%\cr&
       + \frac{1}{4} \catdd  \ln 2
       \biggr)
       \ ,
\end{align}
%%%%%%%%%%%%%%%%%%%%%%%%%%%%%%%%%%%%%%%%%%%%%%%%%%%%%%%%%
\begin{align}
\combel{{51}}=& 
   \intaxonetwos{0,2,0}
 - \frac{9}{4} \intaxonetwos{0,0,1} \ln 2 
       \ ,
\qquad
\combel{{52}}= 
         \intaxonetwos{0,1,1}
       - \frac{3}{8} \intaxonetwos{0,0,2}
       - \frac{3}{2} \intaxonetwos{0,0,1} \ln 2
       \ ,
\\
\combel{{53}}=&
         \intaxoneones{1,0,1}
       - \intaxoneones{0,1,1}
       + \frac{1}{4} \intaxoneones{0,0,2}
       \ ,
\qquad      
\combel{{54}}= 
%        \intaxoneones{0,2,0}
%       - \frac{3}{2} \intaxoneones{0,1,1}
%       + \frac{9}{16} \intaxoneones{0,0,2}
%	   =
           \combel{{51}}-\frac{3}{2}\combel{{52}} 
       \ ,
%%%%\combel{{74}}=&\Z2 \combel{{54}}
\end{align}
\begin{align}
\combel{{61}}=&
          \intaxonetwos{2,1,0}
       + \frac{7}{3} \intaxonetwos{1,2,0}
       - 2 \intaxonetwos{1,1,1}
       + \frac{40}{27} \intaxonetwos{0,3,0}
       - \frac{7}{3} \intaxonetwos{0,2,1}
%\cr&      
       + \intaxonetwos{0,1,2}
       -30 \combel{{54}} \ln 2 
       \ ,
\end{align}
\begin{align}
\combel{{62}}=&
          \intaxonetwos{2,0,1}
       + \frac{14}{3} \intaxonetwos{1,2,0}
       - 2 \intaxonetwos{1,1,1}
       - 2 \intaxonetwos{1,0,2}
       - \frac{370}{27} \intaxonetwos{0,3,0}
\cr&      
       + \frac{85}{3} \intaxonetwos{0,2,1}
       - 22 \intaxonetwos{0,1,2}
       + 7  \intaxonetwos{0,0,3}
       + 11 \Z2 \intaxonetwos{0,0,1}
       -20 \combel{{54}} \ln 2
       \ .
\end{align}
In the above expressions  
 $\zeta(n)=\sum_{i=1}^\oo i^{-n}$,
 $a_n=\sum_{i=1}^\oo 2^{-i}\;i^{-n}$, 
 $b_6=H_{0,0,0,0,1,1}\left(\frac{1}{2}\right)$, 
 $b_7=H_{0,0,0,0,0,1,1}\left(\frac{1}{2}\right)$, 
 $d_7=H_{0,0,0,0,1,-1,-1}(1)$, 
 $\cl{n}{\theta}=\Im \Li_n (e^{i\theta})$.
$C_{8xy}$ are the  $\e^0$ coefficients of the $\e-$expansion of
six master integrals (see Ref.\citep{Laporta:2017okg}).
$H_{i_1,i_2,{\ldots} }(x)$ are the harmonic
 polylogarithms\citep{Remiddi:1999ew,Laporta:2018eos,Ablinger:2011te}.
The integrals $f_j$ are defined as follows:
\begin{eqnarray}\label{fdef}
f_m(i,j,k)&=&\int_1^9 ds\; D_1(s) \Re\left(\sqrt{3^{m-1}} D_m(s)\right)  \left(s-\frac{9}{5}\right)
\ln^i\left(9-s\right)
\ln^j\left(s-1\right)
\ln^k\left(s\right)
\ ,
%\nonumber
\\\cr
%\text{where }
D_m(s)&=&\frac{2}{\sqrt{(\rs+3)(\rs-1)^3}}K\left(m-1-(2m-3)\frac{(\rs-3)(\rs+1)^3}{(\rs+3)(\rs-1)^3}\right)\
;
%\nonumber
\end{eqnarray}
$B_3$ and $C_3$ have hypergeometric
expressions\citep{Laporta:2008sx,Zhou:2018wyp}:
\begin{align}\label{cdef}
%A_3=\int_0^1 dx &\dfrac{K_c(x) K_c(\textup{1-$x$})}{\sqrt{1-x}}= 
%\dfrac{\pi}{54}\sqrt{3}\left(
%{}_4\WF_3\left(\begin{smallmatrix}
%{{\frac{1}{6}\;\frac{1}{3}\;\frac{1}{3}\;\frac{1}{2}}}\\
%{{\frac{5}{6}\;\frac{5}{6}\;\frac{2}{3}}}\end{smallmatrix}; 1\right)
%+
%{}_4\WF_3\left(\begin{smallmatrix}
%{{\frac{5}{6}\;\frac{2}{3}\;\frac{2}{3}\;\frac{1}{2}}}\\
%{{\frac{7}{6}\;\frac{7}{6}\;\frac{4}{3}}}\end{smallmatrix}; 1\right)
%\right)\!\:,\\\cr
B_3=&
\dfrac{\pi}{27}\sqrt{3}\left(
{}_4\WF_3\left(\begin{smallmatrix}
{{\frac{1}{6}\;\frac{1}{3}\;\frac{1}{3}\;\frac{1}{2}}}\\
{{\frac{5}{6}\;\frac{5}{6}\;\frac{2}{3}}}\end{smallmatrix}; 1\right)
-
{}_4\WF_3\left(\begin{smallmatrix}
{{\frac{5}{6}\;\frac{2}{3}\;\frac{2}{3}\;\frac{1}{2}}}\\
{{\frac{7}{6}\;\frac{7}{6}\;\frac{4}{3}}}\end{smallmatrix}; 1\right)
\right) \ , \\\cr
C_3=&
\dfrac{\pi}{27}\sqrt{3}\left(
{}_4\WF_3\left(\begin{smallmatrix}
{{\frac{1}{6}\;\frac{1}{3}\;\frac{4}{3}\;-\frac{1}{2}}}\\
{{-\frac{1}{6}\;\frac{5}{6}\;\frac{5}{3}}}\end{smallmatrix}; 1\right)
-
{}_4\WF_3\left(\begin{smallmatrix}
{{-\frac{7}{6}\;-\frac{1}{3}\;\frac{2}{3}\;-\frac{1}{2}}}\\
{{-\frac{5}{6}\;\frac{1}{6}\;\frac{1}{3}}}\end{smallmatrix}; 1\right)
\right) \ , 
\\\cr
{}_4\WF_3\left(\begin{smallmatrix}
{{a_1\;a_2\;a_3\;a_4}}\\
{{b_1\;b_2\;b_3}}\end{smallmatrix}; x\right)
&
=\dfrac{\Gamma{(a_1)}\Gamma{(a_2)}\Gamma{(a_3)}\Gamma{(a_4)}}{\Gamma{(b_1)}\Gamma{(b_2)}\Gamma{(b_3)}}
{}_4F_3\left(\begin{smallmatrix}
{{a_1\;a_2\;a_3\;a_4}}\\
{{b_1\;b_2\;b_3}}\end{smallmatrix}; x\right)
\ .
\end{align}

\section{ $Z_2$ and $Z_m$ to three loops}
For completeness we list here the analytical expressions 
of $Z_2$ and $Z_m$ at one, two, and three loops,
expanded in $\e$ up to the level needed for five-loop 
renormalization  
(two powers in $\e$ more than the results of 
Ref.\citep{Melnikov:2000zc}
and one power more than Ref.\citep{Marquard:2016dcn,Marquard:2018rwx}).
\begin{align}
Z_2^{(1)}&=Z_m^{(1)}=-\frac{3}{4\e}-\frac{1}{1-2\e}
\ ,
\end{align}
\begin{align}
Z_2^{(2)}& =
            \frac{17}{32 \e^2}
          + \frac{229}{192 \e}
          + \frac{8453}{1152}
          - \frac{3}{2} \Z3
          + 6 \Z2 \ln 2
          - \frac{55}{8} \Z2
       + \e  \biggl(
            \frac{86797}{6912}
          - \frac{419}{16} \Z2
          - \frac{203}{8} \Z3
\cr&
          + \frac{93}{2} \Z2 \ln 2
          + \frac{63}{2} \Z4
          - 12 \Z2 \ln^2 2
          - 24 \comba{{4}}
          \biggr)
       + \e^2   \biggl(
            \frac{2197589}{41472}
          - \frac{3393}{32} \Z2
          - \frac{779}{8} \Z3
          + {180} \Z2 \ln 2
\cr&
          + \frac{1073}{8} \Z4
          - {93} \Z2 \ln^2 2
          - {186} \comba{{4}}
          + \frac{609}{4} \Z5
          + {18} \Z3 \Z2
          - {93} \Z4 \ln 2
          + {36} \Z2 \ln^3 2
          - {144} \comba{{5}}
          \biggr)
\cr&
       + \e^3   \biggl(
            \frac{22329277}{248832}
          - \frac{22387}{64} \Z2
          - \frac{5769}{16} \Z3
          + {594} \Z2 \ln 2
          + {526} \Z4
          - {360} \Z2 \ln^2 2
          - {720} \comba{{4}}
\cr&
          + \frac{14919}{16} \Z5
          + \frac{169}{2} \Z3 \Z2
          - \frac{2883}{4} \Z4 \ln 2
          + {279} \Z2 \ln^3 2
          - {1116} \comba{{5}}
          + \frac{363}{4} \Z6
          - \frac{579}{2} \ZTD
\cr&
          + {300} \Z3 \Z2 \ln 2
          + {324} \Z4 \ln^2 2
          - {54} \Z2 \ln^4 2
          - {720} \comba{{61}}
          + {576} \comba{{62}}
          \biggr)
	   +O(\e^4)
\ ,
\end{align}

\begin{align}
Z_m^{(2)}&=
            \frac{13}{32 \e^2}
          + \frac{73}{64 \e}
          + \frac{475}{128}
          - \frac{3}{4} \Z3
          - \frac{23}{8} \Z2
          + 3 \Z2 \ln 2
       + \e   \biggl(
            \frac{2529}{256}
          - \frac{245}{16} \Z2
          - \frac{47}{4} \Z3
\cr&
          + {24} \Z2 \ln 2
          + \frac{63}{4} \Z4
          - {6} \Z2 \ln^2 2
          - {12} \comba{{4}}
          \biggr)
       + \e^2   \biggl(
            \frac{13379}{512}
          - \frac{1831}{32} \Z2
          - \frac{437}{8} \Z3
          + {96} \Z2 \ln 2
\cr&
          + {80} \Z4
          - {48} \Z2 \ln^2 2
          - {96} \comba{{4}}
          + \frac{609}{8} \Z5
          + {9} \Z3 \Z2
          - \frac{93}{2} \Z4 \ln 2
          + {18} \Z2 \ln^3 2
          - {72} \comba{{5}}
          \biggr)
\cr&
       + \e^3   \biggl(
            \frac{69001}{1024}
          - \frac{12613}{64} \Z2
          - \frac{3115}{16} \Z3
          + {321} \Z2 \ln 2
          + {259} \Z4
          - {192} \Z2 \ln^2 2
          - {384} \comba{{4}}
\cr&
          + \frac{1011}{2} \Z5
          + {49} \Z3 \Z2
          - {372} \Z4 \ln 2
          + {144} \Z2 \ln^3 2
          - {576} \comba{{5}}
          + \frac{363}{8} \Z6
          - \frac{579}{4} \ZTD
\cr&
          + {150} \Z3 \Z2 \ln 2
          + {162} \Z4 \ln^2 2
          - {27} \Z2 \ln^4 2
          - {360} \comba{{61}}
          + {288} \comba{{62}}
          \biggr)
	   +O(\e^4)
\ ,
\end{align}
\begin{align}
 Z_2^{(3)}&=
          - \frac{131}{384 \e^3}
          - \frac{2141}{2304 \e^2}
       + \frac{1}{\e}  \biggl(
          - \frac{116489}{13824}
          + \frac{935}{96} \Z2
          + \frac{17}{8} \Z3
          - \frac{17}{2} \Z2 \ln 2
          \biggr)
       - \frac{2121361}{82944}
          - \frac{197731}{8640} \Z2
\cr&
          + \frac{2803}{144} \Z3
          + \frac{1367}{24} \Z2 \ln 2
          - \frac{383}{8} \Z4
          + {23} \Z2 \ln^2 2
          + {18} \comba{{4}}
          - \frac{5}{16} \Z5
          + \frac{3}{4} \Z3 \Z2
       + \e   \biggl(
          - \frac{200754221}{2488320}
\cr&
          - \frac{14806511}{86400} \Z2
          - \frac{216391}{960} \Z3
          + \frac{5687}{5} \Z2 \ln 2
          - \frac{141}{4} \Z4
          - \frac{8405}{24} \Z2 \ln^2 2
          - \frac{1712}{3} \comba{{4}}
          - \frac{3201}{16} \Z5
\cr&
          + \frac{835}{24} \Z3 \Z2
          + \frac{689}{12} \Z4 \ln 2
          - \frac{193}{3} \Z2 \ln^3 2
          + \frac{388}{3} \comba{{5}}
          - \frac{899}{6} \Z6
          + \frac{29}{32} \ZTD
          + {84} \Z3 \Z2 \ln 2
\cr&
          - {60} \Z4 \ln^2 2
          + {96} \comba{{63}}
          \biggr)
       + \e^2   \biggl(
          - \frac{7232293891}{24883200}
          - \frac{8115586429}{7776000} \Z2
          - \frac{384910567}{259200} \Z3
          + \frac{18690613}{2700} \Z2 \ln 2
\cr&
          + \frac{1946767}{576} \Z4
          - \frac{3203351}{432} \Z2 \ln^2 2
          - \frac{4052957}{540} \comba{{4}}
          + \frac{9351}{2} \Z5
          + \frac{103579}{96} \Z3 \Z2
          - \frac{9511}{4} \Z4 \ln 2
\cr&
          + \frac{24373}{12} \Z2 \ln^3 2
          - \frac{20576}{3} \comba{{5}}
          - \frac{110993}{144} \Z6
          + \frac{30301}{48} \ZTD
          - \frac{1339}{4} \Z3 \Z2 \ln 2
          - \frac{4637}{12} \Z4 \ln^2 2
\cr&
          + \frac{763}{6} \Z2 \ln^4 2
          + {988} \comba{{61}}
          - \frac{2704}{3} \comba{{62}}
          + {62} \comba{{63}}
          + \frac{13801}{32} \Z7
          - \frac{22743}{32} \Z4 \Z3
          + \frac{9695}{32} \Z5 \Z2
\cr&
          - \frac{15435}{32} \Z6 \ln 2
          - {49} \Z3 \Z2 \ln^2 2
          - {98}  \comba{{71}}
          + {392} \comba{{72}}
          + {196} \comba{{4}} \Z3
          + {576} \comba{{5}} \Z2
          + {576} \comba{{73}}
          \biggr)
	   +O(\e^3)
\ ,
\end{align}
\begin{align}
Z_m^{(3)}&=
          - \frac{221}{1152 \e^3}
          - \frac{5561}{6912 \e^2}
       + \frac{1}{\e}   \biggl(
          - \frac{154445}{41472}
          + \frac{391}{96} \Z2
          + \frac{13}{16} \Z3
          - \frac{17}{4} \Z2 \ln 2
          \biggr)
       - \frac{3489365}{248832}
          - \frac{5783}{2880} \Z2
\cr&
          + \frac{719}{72} \Z3
          + \frac{89}{6} \Z2 \ln 2
          - \frac{979}{48} \Z4
          + \frac{65}{6} \Z2 \ln^2 2
          + \frac{23}{3} \comba{{4}}
          + \frac{5}{8} \Z5
          - \frac{3}{8} \Z3 \Z2
       + \e   \biggl(
          - \frac{410529217}{7464960}
\cr&
          - \frac{6911609}{86400} \Z2
          - \frac{240973}{4320} \Z3
          + \frac{41969}{90} \Z2 \ln 2
          - \frac{15215}{288} \Z4
          - \frac{2011}{18} \Z2 \ln^2 2
          - \frac{1756}{9} \comba{{4}}
          - \frac{6985}{96} \Z5
\cr&
          + \frac{91}{8} \Z3 \Z2
          + \frac{51}{8} \Z4 \ln 2
          - \frac{51}{2} \Z2 \ln^3 2
          + {46} \comba{{5}}
          - {25} \Z6
          - \frac{1}{2} \ZTD
          + \frac{63}{4}  \Z3 \Z2 \ln 2
\cr&
          - \frac{45}{4} \Z4 \ln^2 2
          + {18} \comba{{63}}
          \biggr)
       + \e^2   \biggl(
          - \frac{52076602061}{223948800}
          - \frac{1249645817}{2592000} \Z2
          - \frac{69525299}{129600} \Z3
          + \frac{2040043}{675} \Z2 \ln 2
\cr&
          + \frac{553243}{432} \Z4
          - \frac{166889}{54} \Z2 \ln^2 2
          - \frac{397312}{135} \comba{{4}}
          + \frac{109385}{72} \Z5
          + \frac{42119}{96} \Z3 \Z2
          - \frac{3421}{6} \Z4 \ln 2
\cr&
          + {673} \Z2 \ln^3 2
          - {2408} \comba{{5}}
          - \frac{57055}{96} \Z6
          + \frac{2745}{16} \ZTD
          + \frac{235}{4} \Z3 \Z2 \ln 2
          - \frac{509}{4} \Z4 \ln^2 2
          + {230} \comba{{61}}
\cr&
          - {184} \comba{{62}}
          + {150} \comba{{63}}
          + \frac{531}{4} \Z7
          - {447} \Z4 \Z3
          - \frac{219}{2} \Z5 \Z2
          - \frac{945}{8} \Z6 \ln 2
          - {12} \Z3 \Z2 \ln^2 2
\cr&
          + \frac{153}{4} \Z2 \ln^4 2
          - {24}  \comba{{71}}
          + {96}  \comba{{72}}
          + {48}  \comba{{4}} \Z3
          + {516} \comba{{5}} \Z2
          + {516} \comba{{73}}
          \biggr)
	   +O(\e^3)
\ .
\end{align}
%\begin{equation}
%x=\frac{3}{4}+\dfrac{3}{4} +{\tfrac{5}{6}}
%\end{equation}
%%%%%%%%%%%%%%%%%%%%%%%%%%%%%%%%%%%%%%%%%%%%%%%%%%%%%%%%%%%%%%%%%%%%%%%%%%%%%%
\setcounter{secnumdepth}{0} 
\section{Acknowledgments}
This work has been supported by the Supporting TAlent in ReSearch at
Padova University (UniPD STARS Grant 2017 ``Diagrammalgebra'').

I wish to thank Pierpaolo Mastrolia for the encouragement and the support.
I wish to thank Thomas Gehrmann for providing me the access
to the computing facilities of the Institute for Theoretical Physics of Zurich.
%%%%%%%%%%%%%%%%%%%%%%%%%%%%%%%%%%%%%%%%%%%%%%%%%%%%%%%%%%%%%%%%%%%%%%%%%%%%%%
%%%%%%%%%%%%%%%%%%%%%%%%%%%%%%%%%%%%%%%%%%%%%%%%%%%%%%%%%%%%%%%%%%%%%%%%%%%%%%
%\section{References}

%%%%%%%%%%%%%%%%%%%%%%%%%%%%%%%%%%%%%%%%%%%%%%%%%%%%%%%%%%%%%%%%%%%%%%%%
% SCOMMENTARE PER SPEDIRE
%\vfill\eject 
%\pagenumbering{roman}
%\setcounter{page}{1}
%\phantom{.}\vspace{5cm}
%\section*{Figure Captions}
%\par\noindent Figure 1: \CAPTIONFIGA
%\phantom{.}\vspace{12cm}\IDENTIFY
%%%%%%%%%%%%%%%%%%%%%%%%%%%%%%%%%%%%%%%%%%%%%%%%%%%%%%%%%%%%%%%%%%%%%%%%
%\phantom{.}
%\vfill\eject 
%\phantom{.}\vspace{1cm}
%\FigureOne
%\phantom{.}\vspace{1cm}\IDENTIFY
%%%%%%%%%%%%%%%%%%%%%%%%%%%%%%%%%%%%%%%%%%%%%%%%%%%%%%%%%%%%%%%%%%%%%%%%
%%%%%%%%%%%%%%%%%%%%%%%%%%%%%%%%%%%%%%%%%%%%%%%%%%%%%%%%%%%%%%%%%%%%%%%%
%%%%%%%%%%%%%%%%%%%%%%%%%%%%%%%%%%%%%%%%%%%%%%%%%%%%%%%%%%%%%%%%%%%%%%%%
%\vfill\eject 
%\vfill\eject 
%%%%%%%%%%%%%%%%%%%%%%%%%%%%%%%%%%%%%%%%%%%%%%%%%%%%%%%%%%%%%%%%%%%%%%%%%%%%%%
\end{document}